\documentclass[aps,prl,twocolumn,showpacs,amsmath,amssymb,amsfonts,groupedaddress]{revtex4-1}
\usepackage{graphicx}
\begin{document}
\title{Ground State Properties of Spin-Orbit Coupled Bose Gases for Arbitrary Interactions }
\author{Renyuan Liao}\email{rliao08@gmail.com}
\author{Wu-Ming Liu}
\affiliation{National Laboratory for Condensed Matter Physics, Institute of Physics, Chinese Academy of Sciences, Beijing 100190, China}
\date{\today}
\begin{abstract}
We study spin-orbit coupled (SOC) Bose gases with arbitrary interspecies interaction. Besides at a critical interaction our results carry over to the recent publication [PRL 109,025301 (2012)], we identify various new features arising from the interplay of SOC and interspecies interaction, including a roton minimum in the excitation spectrum and dual effects of SOC on ground state energies depending on interspecies interactions. Counterintuitively, we find that at low interspecies interaction the SOC stabilizes the system by suppressing the quantum depletion. We show that the static structure factor is immune to the SOC in the phase space where time-reversal symmetry is preserved.
\end{abstract}
\pacs{67.85.Fg,03.75.Mn,05.30.Jp, 67.85.Jk}
\maketitle
The pioneering experimental realization of synthetic gauge field and spin-orbit coupling (SOC)~\cite{LIN,BLO11,CHE12,ZHA12,ZWI12}
provides fascinating opportunities to explore quantum many-body systems of ultracold atomic gases.
Synthetic gauge potential is defining a new dimension for simulating real materials with cold atoms. The engineered spin-orbit coupling
in a neutral atomic Bose-Einstein condensate was accomplished by dressing two atomic
spin states with a pair of lasers.
In electronic systems, spin-orbit coupling is crucial for quantum spin Hall effects and topological insulators~\cite{XIA},
which has captured a great deal of attention in condensed matter community.
In particular, spin-orbit coupled bosons does not have an analogy in conventional condensed matter systems,
resulting in the emergence of many novel quantum phases  such as striped superfluid phase~\cite{ZHA10,TIN11,YUN12}
and half vortex phase~\cite{WU11,STA08,SIN11,XU11,HU12}.

Inspired by experimental achievements, theorists have been paying vast attention to
SOC Bose gases in many aspects,
including the fluctuations effects~\cite{HUI11,GOL11,TOM12}, the quasiparticle excitation spectrum~\cite{BAR12,XU12},
the depletion of condensate~\cite{BAY12,CUI12} and the correlation functions~\cite{YUN122,GIO12}.
These theoretical works focus mainly on two types of spin-orbit coupling: one is the SOC with
equal Rashba and Dresselhaus couplings firstly realized at NIST~\cite{LIN},
the other is the Rashba SOC which is theoretically more complicated and appealing.
For the Rashba-type SOC, these works~\cite{XU12,BAR12,BAY12} are exclusively dedicated to the critical case where
the intraspecies and interspecies interactions are the same,
leaving the physics of the interplay
between spin-orbit coupling and interspecies interaction, which are both experimentally relevant and
theoretically interesting, largely intact.

In this work, by monitoring the effects of SOC and interspecies interaction, we have identified several new features and clarified several important issues regarding SOC Bose gases. Firstly, by examining the excitation spectrum, we find a roton minimum, a result also found for the SOC with equal Rashba and Dresselhaus coupling~\cite{YUN122}. In addition, we find that by tuning the interspecies interaction to a critical one, a second gapless mode emerges, consistent with Ozawa and Baym's result~\cite{BAY12}. Through inspecting the infrared behaviors around the gapless modes, we are able to conclude that the Bose-Einstein-Condensates (BEC) is stable, in spite of the fact that in the absence of interactions there is no BEC for SOC Bose gases in three dimensions due to constant density of states at low energy.
Secondly, by studying the ground state energy, we find that at low interspecies coupling, the ground state energy actually increases with SOC, in contrast to that of large interspecies coupling. Thirdly, we clarify the effects of SOC on the quantum depletion. At low interspecies coupling, we show that the SOC actually suppresses the quantum depletion, differing from a reported result~\cite{CUI12}.
Finally, we find that the experimentally important quantity, the static structure factor is immune 
to SOC in the parameter space where the time-reversal symmetry is preserved.

We consider three-dimensional homogeneous two-component Bose gases with an isotropic in-plane
Rashba spin-orbit coupling, described by the following grand canonical Hamiltonian:
\begin{eqnarray}
   H&=&\int d^3\mathbf{r}\sum_{\sigma=\uparrow,\downarrow}\left[\psi_\sigma^\dagger\left(\frac{\hbar^2\nabla^2}{2m}-\mu\right)\psi_\sigma+g_1(\psi_\sigma^\dagger\psi_\sigma)^2\right]\nonumber\\
   & &+\int d^3\mathbf{r}\left[2g_{12}\psi_\uparrow^\dagger\psi_\uparrow\psi_\downarrow^\dagger\psi_\downarrow+\left(\psi_\uparrow^\dagger \hat{R}\psi_\downarrow+h.c.\right)\right],
\end{eqnarray}
where $g_1$ is the intraspecies interaction, $g_{12}$ is the interspecies interaction,
and the spin-orbit coupling is $\hat{R}=\lambda(\hat{P_x}-i\hat{P_y})$, with $\lambda$ being the coupling strength.

Within  the framework of imaginary-time field integral, the partition function of the system in
the momentum space can be cast as
$\mathcal{Z}=\int d[\psi_{\mathbf{q}\sigma}^*\psi_{\mathbf{q}\sigma}] \exp{(-S[\psi_{\mathbf{q}\sigma}^*,\psi_{\mathbf{q}\sigma}])}$ with the action
\begin{eqnarray}
S&=&\int d\tau \sum_{\mathbf{q}}\left[\sum_\sigma\psi_{\mathbf{q}\sigma}^*\left(\partial_\tau+\xi_\mathbf{q}\right)\psi_{\mathbf{q}\sigma}+\left(R_\mathbf{q}\psi_{\mathbf{q}\uparrow}^*\psi_{\mathbf{q}\downarrow}+c.c\right)\right]\nonumber\\
& &+\int d\tau \sum_\sigma\sum_{\mathbf{k+l=m+n}}\frac{g_1}{V}\psi_{\mathbf{k}\sigma}^*\psi_{\mathbf{l}\sigma}^*\psi_{\mathbf{m}\sigma}\psi_{\mathbf{n}\sigma}\nonumber\\
& &+\int d\tau \sum_{\mathbf{k+l=m+n}}\frac{2g_{12}}{V}\psi_{\mathbf{k}\uparrow}^*\psi_{\mathbf{l}\downarrow}^*\psi_{\mathbf{m}\downarrow}\psi_{\mathbf{n}\uparrow}.
\end{eqnarray}
Here, we have defined the free particle energy dispersion $\xi_\mathbf{q}=\mathbf{q}^2-\mu$ (we have set $\hbar=2m=k_B=1$)
and $R_\mathbf{q}=\lambda q_\perp e^{-i\varphi_\mathbf{q}}$, with $q_\perp$ being the magnitude of in-plane momentum,
and $\varphi_\mathbf{q}=Arg(q_x,q_y)$. The chemical potential $\mu$ is introduced to fix the total number of the particles,
and $V$ is the volume of the system. For a non-interacting system, the Hamiltonian is diagonalized in helicity
basis with a dispersion $E_\mathbf{q}^{\pm}=\xi_\mathbf{q}\pm\lambda q_\perp$. The lowest energy states are infinitely degenerate,
sitting at the circular ring in the plane $q_z=0$ confined by $q_x^2+q_y^2=\left(\frac{\lambda}{2}\right)^2$ in the momentum space.
Mean-field study~\cite{ZHA10} find that the plane wave (PW) phase exists for $g_1>g_{12}$ and the striped phase exists for $g_1<g_{12}$.
The PW phase is characterized by a condensation at a single momentum state, while the striped phase
is a coherent superposition of two condensates at two opposite momenta.
Here we restrict ourself to studying the PW phase, and further assume that
the condensation occurs at momentum $\vec{\kappa}=(\lambda/2,0,0)$.
We separate a Bose field into a mean-field part and a fluctuating part: $\psi_{\mathbf{q}\sigma}=\phi_{c\sigma}\delta_{\mathbf{q}\vec{\kappa}}+\phi_{\mathbf{q}\sigma}$.
Without loss of generality, the condensate wavefunction in spin space could be chosen as $\Phi(x)=(\phi_{c\uparrow},\phi_{c\downarrow})^T=\sqrt{\frac{n_0}{2}}(1,-1)^T e^{i\kappa x}$,
such that $|\phi_{c\sigma}|=\sqrt{n_{0\sigma}}=\sqrt{n_0/2}$, with $n_0$ being the total condensate density.
Retaining terms of the zeroth and quadratic orders in the fluctuating fields, we rewrite the effective action as $S_{eff}=S_c+S_g$.
Here the mean-field action is $S_c=\beta V\sum_\sigma\left[(\xi_\kappa-\lambda\kappa)n_{c\sigma}+(g_1+g_{12})n_{c\sigma}^2\right]$.
Saddle point conditions $\delta S_c/\delta n_{c\sigma}=0$ leads to $\mu=-\frac{\lambda^2}{4}+(g_1+g_{12})n_0$ and $S_c=-\frac{\beta V}{2}(g_1+g_{12})n_0^2$.
By defining
 a four-dimensional column vector $\Phi_\mathbf{q}=(\phi_{\vec{\kappa}+\mathbf{q}\uparrow},\phi_{\vec{\kappa}+\mathbf{q}\downarrow},\phi_{\vec{\kappa}-\mathbf{q}\uparrow}^*,\phi_{\vec{\kappa}-\mathbf{q}\downarrow}^*)$,
 we can bring the gaussian action into a compact form $S_g=\sum_q\frac{1}{2}\Phi_\mathbf{q}^*\mathcal{G}^{-1}\Phi_\mathbf{q}-\beta\sum_\mathbf{q}\epsilon_{-\mathbf{q}}$,
 where $q=(\mathbf{q},iw_n)$ with $w_n=2\pi n/\beta$ being the bosonic Matsubara frequencies, and the inverse Green's function $\mathcal{G}^{-1}(\mathbf{q},iw_n)$ defined as
\begin{widetext}
\begin{eqnarray}
\mathcal{G}^{-1}(\mathbf{q},iw_n)=
\begin{pmatrix}
-iw_n+\epsilon_\mathbf{q}& R_\mathbf{q}-g_{12}n_0&g_1n_0&-g_{12}n_0\\
R_q^*-g_{12}n_0&-iw_n+\epsilon_\mathbf{q}&-g_{12}n_0&g_1n_0\\
g_1n_0&-g_{12}n_0 & iw_n+\epsilon_{-\mathbf{q}}&R_{-\mathbf{q}}^*-g_{12}n_0\\
-g_{12}n_0&g_1n_0&R_{-\mathbf{q}}-g_{12}n_0& iw_n+\epsilon_{-\mathbf{q}}
\end{pmatrix},
\end{eqnarray}
\end{widetext}
where $\epsilon_\mathbf{q}=\xi_{\vec{\kappa}+\mathbf{q}}+(2g_1+g_{12})n_0=\frac{\lambda^2}{2}+\mathbf{q}^2+\lambda q_x+g_1n_0$ and $R_\mathbf{q}=\lambda(\vec{\kappa}+\mathbf{q})_\perp$.
Throughout our calculation, we will choose $g_1n_0$ as a basic energy scale and $\sqrt{g_1n_0}$ as the corresponding momentum scale. We define a dimensionless parameter $\eta=g_{12}/g_1$ ($\eta\in[0,1]$)
to characterize the strength of interspecies interaction.

The excitation spectrum provides useful insights into a system. It could be obtained by examining the poles of the Green's function. To achieve this, one proceeds by evaluating the determinant of $\mathcal{G}^{-1}(\mathbf{q},iw_n)$,
\begin{eqnarray}
Det[\mathcal{G}^{-1}]&=&(iw_n^2-\omega_{10}^2)\left[(iw_n-2\lambda q_x)^2-\omega_{20}^2\right]-2\lambda^2q_y^2F,\nonumber\\
F&=&iw_n(iw_n-2\lambda q_x)+q^2(q^2+2g_1n_0+\lambda^2)+\nonumber\\
& &(g_1+g_{12})n_0\left[(g_1-g_{12})n_0+\lambda^2\right]-\frac{\lambda^2q_y^2}{2},\label{eq:DetG}
\end{eqnarray}
where $\omega_{10}=\sqrt{q^2\left[q^2+2(g_1+g_{12})n_0\right]}$ and $\omega_{20}=\sqrt{(q^2+\lambda^2)\left[q^2+\lambda^2+2(g_1-g_{12})n_0\right]}$.
By requiring $Det[\mathcal{G}^{-1}(\mathbf{q},iw_n)]=0$, one can obtain the excitation spectrum of the system.
It is easily to verify that $Det\left[\mathcal{G}^{-1}(\mathbf{-q},-iw_n)\right]=Det\left[\mathcal{G}^{-1}(\mathbf{q},iw_n)\right]$,
so that it has two excitation branches.

 In the $q_x$-$q_z$ plane (namely $q_y=0$), it is straightforward to analytically derive two branches of the excitation spectrum from Equation~($\ref{eq:DetG}$):
 $\omega_1=\omega_{10}$ and $\omega_2=2\lambda q_x+\omega_{20}$. For the first branch, there exists a gapless mode at $\mathbf{q}_1=(0,0,0)$.
 Expanding around this gapless point, one finds that the low-energy spectrum is phonon-like collective excitation
 $\omega_1(\delta\mathbf{q})\approx \sqrt{2(g_1+g_{12})n_0(\delta q_x^2+\delta q_z^2)+\mathcal{O}(\delta q_y^4})$. For the second branch, it is always gapped if $g_1>g_{12}$.
 At the critical point $g_1=g_{12}$, a second gapless mode emerges at $\mathbf{q}_2=(-\lambda,0,0)$.
 Expansion around this gapless point leads to a free particle-like anisotropic excitation
 $\omega_2(\mathbf{q}_2+\delta \mathbf{q})\approx\delta q_x^2+\delta q_z^2+g_1n_0/(\lambda^2+2g_1n_0)\delta q_y^2$.
 Judging from the behaviors of the low-energy spectrum around the two gapless modes,
 we conclude that there is no infrared divergence for quantum depletion of the condensate
 in three dimensions at zero temperature as the integral $\int q^2dq\sin{\theta}d\varphi/{w_s(\mathbf{q})} (s=1,2)$ is finite.

\begin{figure}[tbp]
\includegraphics[ width= 0.45\textwidth]{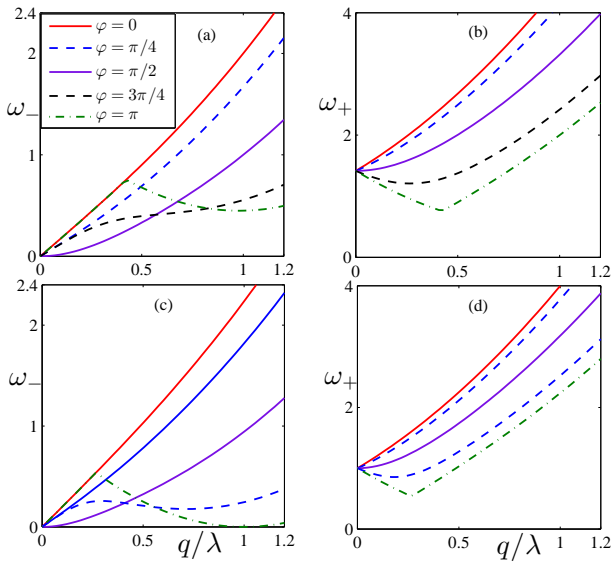}
\caption{(Color online) The excitation spectrum for $\lambda=\sqrt{g_1n_0}$ at zero temperature: (a) the lower branch $\omega_-$ for $\eta=0.5$, (b) the upper branch $\omega_+$ for $\eta=0.5$, (c) the lower branch $\omega_-$ for $\eta=1.0$, and (d) the upper branch $\omega_+$ for $\eta=1.0$. Here $\eta=g_{12}/g_1$.}
\label{Fig1}
\end{figure}

 The behavior of these two branches of the excitation spectrum are summarized in Fig.~\ref{Fig1}.
 As the spectrum enjoys the inversion symmetry along $x$-axis: $\omega_{\pm}(q_x,q_y,0)=\omega_\pm(q_x,-q_y,0)$, we only plot the spectrum
 for the azimuthal angle $\varphi$ lying in $[0,\pi]$. For intermediate interspecies interaction $\eta=0.5$, we plot the two branches in panel (a) and (b).
 The lower branch $\omega_-$ (Fig.~\ref{Fig1}a) accommodates gapless excitation at zero momentum along all directions.
 For $\varphi=0$, $\pi/4$, $\pi/2$, and $3\pi/4$, the quasiparticle energy for both branches increases monotonically with the momentum.
 Remarkably, for $\varphi=\pi$, there appears a roton minimum, signaling that the system has the tendency toward crystallization.
 For the critical interspecies interaction $\eta=1$, we plot the two branches in panel (c) and (d).
 For the lower branch, the quasiparticle energy for both branches increases with the momentum for $\varphi=0$, $\pi/4$, $\pi/2$.
 For $\varphi=3\pi/4$, both branches have a local minimum at  certain finite momentum. At $\varphi=\pi$,
 a second gapless excitation develops at $\mathbf{q}=(-\lambda,0,0)$.

The thermodynamic potential is given by $\Omega=-\ln{\mathcal{Z}}/\beta=\Omega_c+\Omega_g$, where $\Omega_c=-V(g_1+g_{12})n_0^2/2$ and
$\Omega_g=\frac{\beta}{2}Tr\ln{\mathcal{G}^{-1}}-\sum_{\mathbf{q}}\epsilon_\mathbf{q}$.
\begin{figure}[tbp]
\includegraphics[width=0.45\textwidth]{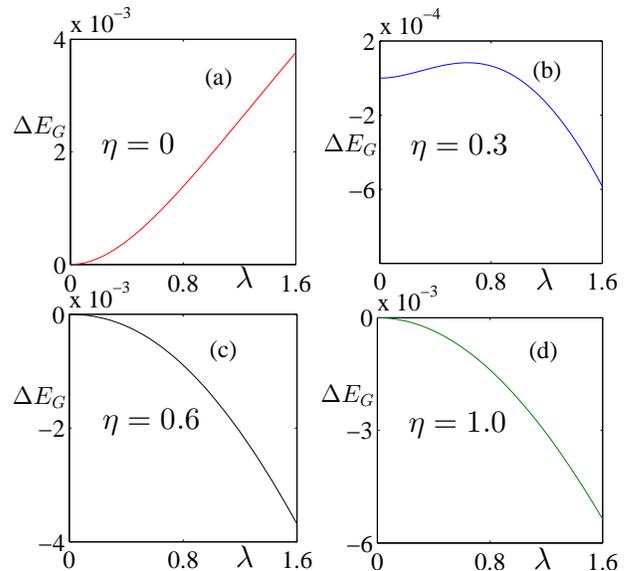}
\caption{(Color online) The shift of the ground state energy $\Delta E_G=E_G(\lambda)-E_G(\lambda=0)$ [in units of $V(g_1n_0)^{5/2}$] as
a function of spin-orbit coupling strength $\lambda$ (in units of $\sqrt{g_1n_0}$) for different interspecies interaction parameter $\eta$: (a) $\eta$=0, (b) $\eta=0.3$, (c) $\eta=0.6$, and (d) $\eta=1.0$. Here $\eta=g_{12}/g_1$.}
\label{Fig2}
\end{figure}
The thermodynamic potential $\Omega$ possesses an ultraviolet divergence, an artifact of zero-range interactions, which can be removed
either by replacing the bare interactions $g_1$ and $g_{12}$ with $T$-matrix or by subtracting counter terms~\cite{AND04}.
At zero temperature, the ground state energy  becomes $E_G=\Omega+\mu N$, renormalized as
\begin{eqnarray}
E_G=E_{MF}+\sum_{\mathbf{q}s}\left[\frac{\omega_s(\mathbf{q})-\epsilon_{\mathbf{q}}}{2}+\frac{(g_1^2+g_{12}^2)n_0^2}{2q^2}\right].
\end{eqnarray}
 Here $E_{MF}=V(g_1+g_{12})n_0^2/2$ is the mean-field energy, independent of the SOC strength $\lambda$.
 The effects of spin-orbit coupling on the shift of the ground state energy $\Delta E_G=E_G(\lambda)-E_G(\lambda=0)$ is shown in Fig.~\ref{Fig2}.
In the absence of interspecies interaction (Fig.~\ref{Fig2}a), the ground state energy increases with the SOC strength $\lambda$.
At $\eta=0.3$, $\Delta E_G$  increases with SOC when SOC is small,
and decreases with SOC when SOC is large. When $\eta$ is sufficiently large (Fig.~$\ref{Fig2}$c,d), $\Delta E_G$ decreases with SOC monotonically.
In the absence of the interspecies interaction and SOC,
we have verified that the ground state energy for either species recovers the Lee-Huang-Yang result~\cite{LHY57} for
spinless weakly-interacting Bose gases with $E_G/V=\frac{\mu n}{2}(1+\frac{128}{15\sqrt{\pi}})\sqrt{na^3}$, where $a$ is the scattering length.
Up to the gaussian level, the ground state energy is universal in the sense that it depends solely on the gas parameter $na^3$
and not on microscopic details of the interaction potential.

\begin{figure}[tbp]
\includegraphics[width=0.45\textwidth]{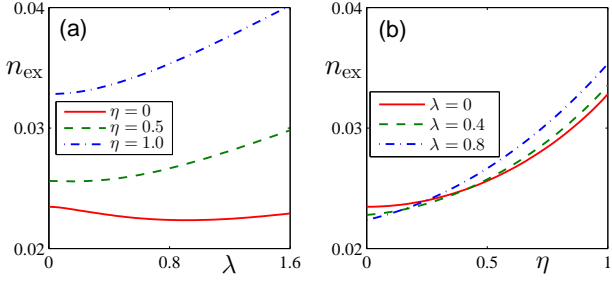}
\caption{(Color online) The number of excited particles $n_{\mathrm{ex}}$ [in units of $(g_1n_0)^{3/2}$] as a function of (a) spin-orbit coupling strength $\lambda$
  (in units of $\sqrt{g_1n_0}$) for different interspecies interaction parameter $\eta$, and (b) interspecies coupling parameter $\eta$ for different spin-orbit coupling strength $\lambda$.}
\label{Fig3}
\end{figure}

Being an intrinsic property of a BEC, the quantum depletion provides key information about the robustness of the superfluid state.
The number of excited particles is given by
 \begin{eqnarray}
 n_{\mathrm{ex}}=2\sum_{\mathbf{q},iw_n} G_{11}(\mathbf{q},iw_n).
 \end{eqnarray}
  We show the quantum depletion of the condensate $n_{\mathrm{ex}}$ in Fig.~\ref{Fig3}. In the absence of interspecies interaction ($\eta=0$, see Fig.~$\ref{Fig3}$a),
$n_{\mathrm{ex}}$ develops a global minimum at a critical SOC.
At sufficiently large interspecies interaction ($\eta$=0.5, 1), $n_{\mathrm{ex}}$ increases monotonically with $\lambda$,
leading to an enhanced quantum depletion.
The effect of interspecies interaction on $n_{\mathrm{ex}}$ is shown in panel (b) for three typical spin-orbit coupling strengths $\lambda$=0, 0.4 and 0.8.
For a fixed SOC strength , the number density of the excited particles increases monotonically with interspecies coupling parameter $\eta$. At low interspecies
interaction, the SOC and interspecies interaction counteract on the quantum depletion; while at sufficiently large interspecies coupling, they work cooperatively.
Being proportional to $(g_1n_0)^{3/2}$, $n_{\mathrm{ex}}$ is small when Bogoliubov condition $g_1n_0\ll1$ is satisfied, justifying our treatment.

\begin{figure}[tbp]
\includegraphics[width=0.45\textwidth]{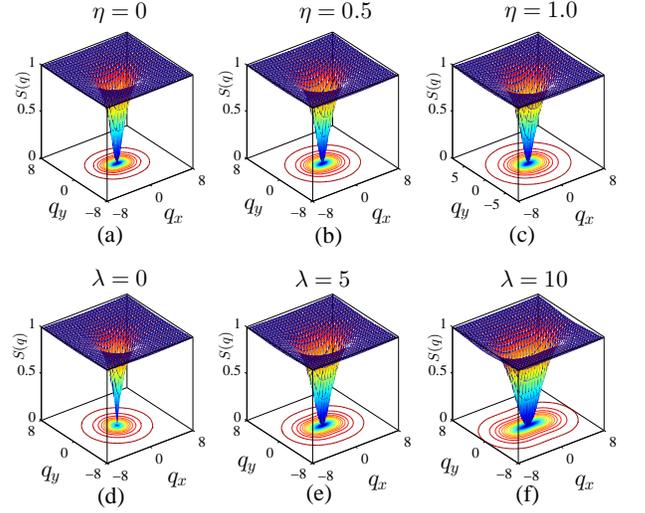}
\caption{(Color online) The static structure factor $S(q)$ as a function of in-plane momentum $\mathbf{q}=(q_x,q_y,0)$.
Upper panel: at $\lambda=3$ (in units of $\sqrt{g_1n_0}$) for different interspecies coupling strength (a) $\eta=0$, (b) $\eta=0.5$,
and (c) $\eta=1.0$. Lower panel: at $\eta=0.5$ for different spin-orbit coupling strength $\lambda$ (d) $\lambda=0$, (e) $\lambda=4$,
and (f) $\lambda=8$. The contours on the bottom plane is the projection of $S(q)$, presented for aesthetic appeal.}
\label{Fig4}
\end{figure}

The static structure $S(\mathbf{q})$ probes  density fluctuations of a system. It provides information on both the spectrum of collective excitations, which could be
investigated at low momentum transfer, and the momentum distribution, which characterizes the behavior of the system at high momentum transfer, where the response is
dominated by single-particle effects. We can evaluate the static structure factor at the Bogoliubov level as follows:
\begin{eqnarray}
NS(\mathbf{q})&=&<\delta \rho_\mathbf{q}^\dagger\delta\rho_\mathbf{q}>\nonumber\\
&=&\frac{N_0}{2}\sum_{iw_n}\sum_{i,j=1}^4(-1)^{i+j}G_{ij}\nonumber\\
&=&N_0\sum_{iw_n}\frac{(-2q^2)A(\mathbf{q},iw_n)}{Det(\mathcal{G}^{-1})},
\end{eqnarray}
where $A(\mathbf{q},iw_n)=(iw_n-2\lambda q_x)^2-(q^2+\lambda^2-\lambda^2q_y^2/q^2)[q^2+\lambda^2+2(g_1-g_{12})n_0]$.
In the $q_x-q_z$ plane, namely $q_y=0$, we have
\begin{eqnarray}
S(q_x,q_z)
=\frac{N_0}{N}\frac{q^2}{\omega_{10}(q)}\coth{\frac{\beta\omega_{10}(q)}{2}}.\label{eq:Sq}
\end{eqnarray}
This is exactly the Feynman relation~\cite{FEY54},
which connects the static structure factor to the excitation spectrum for a Bose system with time-reversal symmetry.
Therefore for $q_y=0$, we find that the Feynman relation is preserved. For any other direction, numerical results suggest that the Feynman relation is violated.
This is reasonable as the ground state breaks time-reversal symmetry except in the $q_x$-$q_z$ plane.
$S(\mathbf{q})$ possesses the inversion symmetry $S(-\mathbf{q})=S(\mathbf{q})$, and increases with the magnitude of the momentum, approaching unit at high momentum.
The static structure factor $S(\mathbf{q})$ in terms of in-plane momentum (we choose $q_z=0$) is shown in Fig.~\ref{Fig4}.
For the upper panel, we plot $S(\mathbf{q})$ at  $\lambda=3$ (in units of $\sqrt{g_1n_0}$) for different interspecies interaction ($\eta$=0, 0.5, 1).
$S(\mathbf{q})$ increases monotonically with in-plane momentum but decreases with $\eta$. The interspecies interaction $
\eta$ has similar effects on $S(\mathbf{q})$ along $q_x$-axis and $q_y$-axis.
In the lower panel, we plot $S(\mathbf{q})$ at  $\eta=0.5$ for different SOC ($\lambda$=0, 4, 8).
 $S(\mathbf{q})$ along $y$-axis decreases with SOC, in contrast to the case along $q_x$-axis, where $S(\mathbf{q})$ remains the same.
 This could be explained from Equation~($\ref{eq:Sq}$) as $S(\mathbf{q})$ does not depend on $\lambda$ for $q_y=0$.

Our predictions bear consequences for experimental observation.
The anisotropic nature of the excitation spectrum could be probed by momentum-resolved photoemission  spectroscopy~\cite{STE08}.
Via measuring in situ density distribution, one can determine the ground state energy of the system~\cite{SAL11}. The depletion of the condensate could be directly
observed as a diffuse background in the time-of-flight images~\cite{KET06}.  The Bragg spectroscopy can be employed to measure
the static structure factor of the system~\cite{KET99}.

In summary, we have studied spin-orbit coupled Bose gases with intraspecies and interspecies interactions.
We have identified various new features arising from the interplay of the SOC and interspecies interactions.
We hope that our work will add new excitement to the surging field of spin-orbit coupled quantum gases.

R. Liao acknowledge helpful discussions with Sandy Fetter, Christophe Salomon and Jason Ho. The work has been supported by
NBRPC under Grant No. 2011CBA00200, NKBRSFC under grants Nos. 2011CB921502, 2012CB821305 and NSFC under Grant No. 11274064, 10934010.

\end{document}